\documentstyle[pre,aps]{revtex}

\newcommand{\be}{\begin{equation}}
\newcommand{\ee}{\end{equation}}
\newcommand{\vp}{\varphi}
\newcommand{\ra}{\rightarrow}
\newcommand{\brf}{\stackrel{-}{f}}
\newcommand{\brM}{\stackrel{-}{M}}

\begin{document}

\draft

\author{S. Gluzman$^1$ and V. I. Yukalov$^2$\footnote{The author to whom 
correspondence is to be addressed}}

\address{$^1$International Center of Condensed Matter Physics\\
University of Brasilia, CP 04513, Brasilia, DF 70919-970, Brazil \\
and \\
$^2$Bogolubov Laboratory of Theoretical Physics \\
Joint Institute for Nuclear Research, Dubna 141980, Russia}

\title{Renormalization Group Analysis of October Market Crashes}

\maketitle

\begin{abstract}

The self--similar analysis of time series, suggested earlier by the authors,
is applied to the description of market crises. The main attention is payed to
the {\it October 1929, 1987} and {\it 1997} stock market crises, which can be
successfully treated by the suggested approach. The analogy between market
crashes and critical phenomena is emphasized.
\end{abstract}

\vspace{1cm}


\section{Self--Similar Analysis}

Renormalization group approach is known to be a powerful tool for treating
critical phenomena in statistical physics. An interesting example of complex
statistical systems are markets [1-5], and market crashes are somewhat
analogous to critical phenomena [1,6]. Keeping in mind this analogy, we have
recently proposed [7] that the time series describing stock--market crises
can be treated by means of resummation or renormalization methods of
theoretical physics. The method we suggested [7] is based on the algebraic
self--similar renormalization [8-10] which is a specific variant of the
self--similar approximation theory [11-15]. The self--similar analysis 
developed in our previous paper [7] for treating stock--market crises was 
shown to describe well a number of such crises from the past. Also, we 
attempted to predict the behaviour of some stock--market indices at the 
end of {\it October 1997}, before the so--called market correction 
occurred. Thus, for the NYSE Composite index we predicted the value
$478.855$. Now we know that the actual value of this index on {\it October
31} was $481.14$. So, the error of our forecast is only $-0.47\%$. For the
Standard and Poor $500$ index we found the value $935.082$, while its actual
value on {\it October 31} was $914.62$. Consequently, the error is $2.24\%$.
And for the Dow Jones index we predicted the value $7788$, which, as compared
to the realized value $7442.07$, makes the error of $4.65\%$. This shows that
we correctly predicted the fall of these stock--market indices before it
actually occurred at the end of {\it October 1997}.

In the present paper we use the self--similar analysis [7] for considering in
detail the famous {\it October 1929} and {\it October 1987} stock market 
crashes as well as the {\it October 1997} crisis. The general scheme of 
the method has been thoroughly described in Ref. [7], because of which we 
shall not repeat it here, in full, but, for convenience, we will remind 
the main steps necessary for the analysis.

Assume that we are considering a function of time, $f(t)$, which characterizes
the market activity. For instance, $f(t)$ can be the price of some security or
commodity, or it can be some price index. Let the values of $f(t)$ be known
for $n$ equidistant successive moments of time, $t=0,1,2,\ldots,n-1$,
preceding a crash,
\be
f(0) = a_0, \qquad f(1)=a_1, \qquad \ldots , \qquad  f(n-1)=a_{n-1} .
\ee
Our aim is to find $f(n)$ at the time $t=n$. The set of data (1) can be
presented in the form of the polynomial
\be
f(t) =\sum_{k=0}^{n-1} A_kt^k \qquad ( 0\leq t\leq n-1) ,
\ee
with the coefficients $A_k$ to be determined from the set of equations (1).
The polynomial representation (2) means that there are the following $n$
approximations for the function $f(t)$ considered:
\be
p_0(t) =A_0=a_0, \quad p_1(t) =p_0(t)+A_1t, \quad \ldots, \quad
p_{n-1}(t)=p_{n-2}(t)+A_{n-1}t^{n-1} .
\ee
For the sequence $\{ p_k(t)\}$, we construct the dynamical system called the
approximation cascade, $\{ y_k\}$, whose trajectory $\{ y_k(\vp,s)\}$ consists
of the points $y_k(\vp,s)\equiv P_k(t(\vp,s),s)$, such that
$P_k(t,s)\equiv t^sp_k(t)$ and the function $t(\vp,s)=(\vp/a_0)^{1/s}$ 
is defined by the equation $P_0(t,s)=a_0t^s=\vp$.
The evolution equation for the approximation cascade $\{ y_k\}$ can be
written in the functional form, $y_{k+p}(\vp,s)=y_k(y_p(\vp,s),s)$, or in
the integral form yielding
$$ \int_{P_{k-1}}^{P_k^*} \frac{d\vp}{v_k(\vp,s)} = \tau , \qquad
v_k(\vp,s) = y_k(\vp,s) - y_{k-1}(\vp,s) , $$
where $P_k=P_k(t,s)$ and $P_k^*=P_k^*(t,s,\tau)$ is a quasifixed point, with
$\tau$ being the minimal time necessary for reaching the quasifixed point
$P_k^*$. Finding the latter and accomplishing the transform
$\lim_{s\ra\infty}t^{-s}P_k^*(t,s,\tau)$ for each $k=1,2,\ldots,n-1$, as is
prescribed by the self--similar bootstrap [10], we come to the sequence of
the self--similar exponential approximants
\be
f_k^*(t,\tau)=A_0\exp\left ( \frac{A_1}{A_0}t\exp\left (\frac{A_2}{A_1}t
\ldots\exp\left (\frac{A_k}{A_{k-1}}\tau t\right )\right )\ldots\right ) .
\ee

The local stability, or the local convergence [15], of the sequence
$\{ f_k^*(t,\tau)\}$ is characterized by the local multipliers
\be
M_k(t,\tau)\equiv\frac{\delta f_k^*(t,\tau)}{\delta f_1^*(t,1)} .
\ee
The practical way of calculating the latter is as follows: From the equation
$f_1^*(t,1)=\vp$, we define the function $t(\vp)=(A_0/A_1)\ln(\vp/A_0)$;
then, introducing $z_k(\vp,\tau)\equiv f_k^*(t(\vp),\tau)$, we may write
$$ M_k(t,\tau)=\left [\frac{\partial z_k(\vp,\tau)}{\partial\vp}
\right ]_{\vp=f_1^*(t,1)} \; . $$
Am important particular case is when $\tau=1$, giving the multiplier
\be
M_k(t) \equiv M_k(t,1) .
\ee

If $|M_k(n)|<1$, then the sequence $\{ f_k^*(t,1)\}$ is locally stable at
$t=n$, and $f_{n-1}^*(n,1)$ is to be a reasonable approximation for the
function $f(t)$ at $t=n$. When the terms $f_{n-1}^*(n,1)$ and
$f_{n-2}^*(n,1)$ are noticeably different from each other, this means that
we are yet far from the fixed point. One possibility then could be to define a
Cesaro average of the corresponding approximations [14,15]. Another option is
to locate the quasifixed point by imposing the minimal--difference condition
\be
| f_{n-1}^*(t,\tau_n) - f_{n-2}^*(t,\tau_n)| =
\min_{\tau} | f_{n-1}^*(t,\tau) - f_{n-2}^*(t,\tau)| ,
\ee
which defines the corresponding effective time $\tau_n=\tau_n(t)$. The
simplest variant of condition (7) is the equality
$f_{n-1}^*(t,\tau)=f_{n-2}^*(t,\tau)$ resulting in the equation
\be
\tau =\exp\left (\frac{A_{n-1}}{A_{n-2}}t\tau\right ) .
\ee
Defining $\tau=\tau_n(t)$ from (8) and substituting it into 
$f_{n-1}^*(t,\tau)$, we obtain a forecast for the time $t\geq n$,
\be
f_{n-1}^*(t)\equiv f_{n-1}^*(t,\tau_n(t)) .
\ee
The stability of the fixed point (9) is characterized by the multiplier
\be
M_{n-1}^*(t)\equiv \frac{1}{2}\left [ M_{n-1}(t,\tau_n(t)) + 
M_{n-2}(t,\tau_n(t))\right ] .
\ee

Another way of defining a quasifixed point is through the average
\be
\brf_{n-1}(t) \equiv\frac{1}{2}\left [ f_{n-1}^*(t,1) + f_{n-2}^*(t,1)
\right ] .
\ee
This definition is valid even when Eq. (8) has no solution. The 
multiplier, characterizing the stability of the quasifixed point (11), 
can be defined as
\be
\brM_{n-1}(t) \equiv\frac{1}{2}\left [ M_{n-1}(t,1) + 
M_{n-2}(t,1)\right ] .
\ee

Varying the number $n=3,4,5,\ldots$, we obtain a set of possible 
forecasts (9) and (11). The {\it optimal forecast} is, by definition, that
corresponding to the minimal absolute value of a multiplier from the
family of all available multipliers (10) and (12).

It is worth emphasizing that our main idea of treating market dynamics in 
the vicinity of a crisis as a self--similar evolution is based on the 
analogy between market crises and critical phenomena. The collective crowd
behaviour of many interacting agents becomes prevailing near a market crisis
[6,16]. In the precrisis region, the market dynamics can be described as 
a superposition of two types of temporal modes. One is the dominant slow 
mode corresponding to the collective behaviour, and all others are fast 
modes caused by random individual interactions and external sources. In 
other words, the dominant collective mode describes the coherent 
behaviour of strongly correlated market agents, while the subordinated 
fast modes correspond to the stochastic incoherent motion of these 
agents. The development of such a coherent behaviour is a necessary 
condition for the formation of a law of collective motion, which, in 
turn, can be expressed as a self--similar evolution. It may happen that 
among the subordinated fast modes there is a hierarchy, so that the 
slowest among these fast modes, being influenced by the collective 
behaviour, displays, in the precrisis region, specific features. This,
for instance, can have to do with the appearance of the log--periodic 
oscillations [1,6,17] near financial crashes and near the crisis phenomena
in several other systems [18]. Such precursor phenomena are also similar 
to heterophase fluctuations occurring in statistical systems near phase 
transitions [19].

In this way, although the property of a market at each time moment is, in 
general, related to all its previous history [20], but in the vicinity of 
a crisis, there appears a principally new feature -- {\it collective coherent
behaviour}. It is just this collective behaviour that makes it possible 
to formulate, by means of the self--similar analysis, a law of motion for 
a market and to forecast crises. And also, it is because of this coherent 
behaviour, an accurate description of a crash may be achieved with the data
for only a few temporal points preceding the crash.

\section{October Crashes}

Now we pass to the application of the method to the series of the New 
York Stock Exchange (NYSE) Composite index in the course of the {\it October
1987} and {\it October 1997} crisis, when the index changed sharply 
during the time comparable to the resolution of the time series. The 
choice of the NYSE Composite index is caused by the easy availability of 
the data stored in the NYSE Historical Statistical Archive in Internet. 
For the {\it October 1929} crash the NYSE Composite index is not 
available, because of which we consider the time series for the Standard 
Statistic index.

\vspace{5mm}

{\large{\bf A. October 1997 Crisis}}

\vspace{2mm}

Considering the corresponding events, we make the self--similar analysis
for different number of points. Below, we describe this analysis for the 
time series of the NYSE Composite index with one month resolution, aiming 
to forecast the value of the index for {\it October 31, 1997}.

\vspace{2mm}

{\bf Three--point analysis}. The following historical data are available:
$$ a_0=494.50\; (July\; 31,\; 97), \quad a_1=470.48, \quad
a_2=497.23\; (Sept.\; 30,\; 97) . $$
From condition (1), the coefficients of polynomial (2) are $A_0=a_0,\; 
A_1=-49,405$, and $A_2=25.385$. For the exponential approximant (4) at 
$t=3$ and $\tau=1$, we have $f_1^*(3,1)=366.435$ and 
$f_2^*(3,1)=463.768$. The corresponding multipliers defined in (6) are
$M_1(3)=1$ and $M_2(3)=-0.147$. The minimal--difference condition (7), 
with $n=3$ and $t=3$, gives the effective time $\tau=\tau_3(3)=0.4784$. 
Then, the self--similar exponential approximant (9) becomes 
$f_2^*(3)=428.447$, and the multiplier (10) is $M_2^*(3)=0.332$. The 
averages (11) and (12) are $\brf_2(3)=415.102$ and $\brM_2(3)=0.427$, 
respectively.

\vspace{2mm}

{\bf Four--point analysis}. The dynamics of the considered index from 
June 30, 1997 to September 30, 1997 is given by the data
$$ a_0=462.44\;\; (June\; 30,\; 97), \quad a_1=494.50, \quad a_2=470.48
\quad a_3=497.23\;\; (Sept.\; 30,\; 97) . $$
The polynomial coefficients of polynomial (2) are $A_1=95.717,\; A_2=-81.465$,
and $A_3=17.808$. For approximants (4) we get $f_2^*(4,1)=475.338$ and 
$f_3^*(4,1)=564.892$, with the related multipliers $M_2(4)=0.106$ and 
$M_3(4)=-0.036$. From the minimal--difference condition (8), at $t=n=4$, 
we find $\tau=\tau_4(4)=0.5946$. Thus, the self--similar exponential 
approximant (9) yields $f_3^*(4)=515.886$, while multiplier (10) gives 
$M_3^*(4)=0.032$. For the averages (11) and (12) we have $\brf_3(4)=520.115$ 
and $\brM_3(4)=0.035$. 

\vspace{2mm}

{\bf Five--point analysis}. The corresponding data are
$$ a_0=441.78\;\; (May\; 30,\; 97), \qquad a_1=462.44, \qquad a_2=494.50 , $$
$$ a_3=470.48, \qquad a_4=497.23 \;\; (Sept.\; 30,\; 97) . $$
From condition (1), we find the coefficients of polynomial (2), 
$A_1=-51.116,\; A_2=119.341,\; A_3=-54.829$, and $A_4=7.264$. Formula (4) 
gives $f_3^*(5,1)=369.416$ and $f_4^*(5,1)=434.65$. The mapping 
multipliers (6) are $M_3(5)=1.163$ and $M_4(5)=-0.056$. The 
minimal--difference condition (8) results in $\tau=0.6501$. From (9) we 
have $f_4^*(5)=423.595$, multiplier (10) being $M_4^*(5)=0.18$. The 
averages (11) and (12) are $\brf_4(5)=402.033$ and $\brM_4(5)=0.554$.

\vspace{2mm}

{\bf Six--point analysis}. In the same way, from the data
$$ a_0=416.94\;\; (Apr.\; 30,\; 97), \qquad a_1=441.78, \qquad a_2=462.44, $$
$$ a_3=494.50, \qquad a_4=470.48, \qquad a_5=497.23\;\; (Sept.\; 30,\; 
97) , $$
we find the polynomial coefficients $A_1=104.366,\; A_2=-155.195,\; 
A_3=98.434,\; A_4=-24.91$, and $A_5=2.145$. Following the standard 
prescription, we get $f_4^*(6,1)=430.124$ and $f_5^*(6,1)=520.109$, with 
the corresponding multipliers $M_4(6)=-0.022$ and $M_5(6)=0.03$. The 
minimal--difference condition (8) gives $\tau=0.6974$. Thence, for 
approximant (9) and multiplier (10), we obtain $f_5^*(6)=478.855$ and 
$M_5^*(6)=0.023$, while for (11) and (12), we find $\brf_5(6)=475.117$ 
and $\brM_5(6)=0.004$, respectively.

As is explained in the first section of the paper, the optimal forecast 
is that corresponding to the minimal modulus of the related multiplier, 
which means that the found fixed point is the most stable one. In the 
above case, the optimal forecast is that given by the six--point 
analysis, $\brf_5(6)=475.117$, which, compared to the {\it October 31, 
1997} index $484.14$, has the error $-1.25\%$.

\vspace{5mm}

{\large{\bf B. October 1987 Crash}}

\vspace{2mm}

Now we consider the behaviour of the NYSE Composite index before the 
October 1987 crash. The data are taken with the three month resolution. 
Our aim is to make a forecast for {\it October 30, 1987}.

\vspace{2mm}

{\bf Three-point analysis}. The data for the considered index from the 
first quarter to the third quarter of 1987 are
$$ a_0=156.11\;\; (Jan.\; 30,\; 87), \qquad a_1=162.86, \qquad 
a_2=178.64\;\; (July\; 31,\; 87) . $$
For the polynomial coefficients, we get $A_1=2.235$ and $A_2=4.515$. The 
sequence of exponential approximants is locally unstable, since 
$M_1(3)=1$ and $M_2(3)\sim 10^{11}$. As an estimate, the value 
$f_1^*(3,1)=162.961$ can be taken.

\vspace{2mm}

{\bf Four--point analysis}. From the fourth quarter of 1986 to the third 
quarter of 1987, the data are
$$ a_0=140.42\;\; (Oct.\; 31,\; 86), \quad a_1=156.11, \quad a_2=162.86,
\quad a_3=178.64\;\; (July\; 31,\; 87) . $$
For the polynomial coefficients, we have $A_1=26.15,\; A_2=-13.455$, and 
$A_3=2.995$. Repeating the standard steps, we find $f_2^*(4,1)=154.433$ 
and $f_3^*(4,1)=193.381$. The corresponding multipliers are $M_2(4)=-0.071$ 
and $M_3(4)=0.255$. From the minimal--difference condition, we get 
$\tau=0.591$. Finally, the self--similar exponential approximant (9) is 
$f_3^*(4)=175.109$, with the multiplier $M_3^*(4)=0.018$, while the average
(11) becomes $\brf_3(4)=173.907$, with the multiplier $\brM_3(4)=0.092$.

\vspace{2mm}

{\bf Five--point analysis}. From the data
$$ a_0=135.89\;\; (July\; 31,\; 86), \qquad a_1=140.42, \qquad 
a_2=156.11, $$
$$ a_3=162.86, \qquad a_4=178.64 \;\; (July\; 31,\; 87) , $$
we get the polynomial coefficients $A_1=-17.268,\; A_2=33.079,\; A_3=-12.868$,
and $A_4=1.586$. Then we find $f_3^*(5,1)=115.623$ and $f_4^*(5,1)=132.899$, 
with $M_3(5)=0.937$ and $M_4(5)=-0.065$. The minimal--difference condition 
gives $\tau=0.664$. The resulting self--similar exponential approximant (9)
is $f_4^*(5)=129.821$, with the multiplier $M_4^*(5)=0.139$, and the average 
(11) is $\brf_4(5)=124.261$, with the multiplier $\brM_4(5)=0.436$.

\vspace{2mm}

{\bf Six--point analysis}. Being based on the data
$$ a_0=135.75\;\; (Apr.\; 30,\; 86), \qquad a_1=135.89, \qquad
a_2=140.42 , $$
$$ a_3=156.11, \qquad a_4=162.86, \qquad a_5=178.64\;\; (July\; 31, \; 86) , $$
we have the polynomial coefficients $A_1=19.907,\; A_2=-40.564,\; 
A_3=26.787,\; A_4=-6.531$, and $A_5=0.541$. In the usual way, we get 
$f_4^*(6,1)=136.656$ and $f_5^*(6,1)=147.008$, the related multipliers 
being $M_4(6)=-0.019$ and $M_5(6)=0.031$. The minimal--difference condition
yields $\tau=0.7045$. The self--similar exponential approximation (9) is
$f_5^*(6)=141.991$, with the multiplier $M_5^*(6)=0.03$, while for (11) 
we get $\brf_5(6)=141.832$, with the multiplier $\brM_5(6)=0.006$.

Among all multipliers from the sets $\{ M_{n-1}^*(n)\}$ and 
$\{\brM_{n-1}(n)\}$, with $n=3,4,5,6$, the multiplier $\brM_5(6)$ has the 
minimal absolute value. Therefore, as the optimal forecast, we accept 
$\brf_5(6)=141.832$. The actual value of the index on {\it October 30, 
1987} was $140.8$, so that our forecast differs from it only by $0.73\%$.

\vspace{5mm}

{\large{\bf C. October 1929 Crash}}

\vspace{2mm}

The historical data from the League of Nations Statistical Yearbook for 
the Standard Statistics index of the New York stock market from April 
1929 to September 1929, with one month resolution are
$$ 193\; (Apr.), \quad 193\; (May), \quad 191\; (June), \quad
203\; (July), \quad 210\; (Aug.), \quad 216\; (Sept.) , $$
where the value for 1926 is taken for 100.

Similarly to the cases expounded above, we find in the three--point analysis
$f_2^*(3)=222.921$ and $M_2^*(3)=0.761$, in the four--point analysis, we 
get $f_3^*(4)=222.916$ and $M_3^*(4)=0.152$, and in the five--point 
analysis, we have $f_4^*(5)=179.503$ and $M_4^*(5)=0.175$. We present the
six--point analysis in more details. In the latter case, the polynomial 
coefficients are $A_1=26.683,\; A_2=-49.208,\; A_3=28.333,\; A_4=-6.292$, 
and $A_5=0.483$. For the exponential approximants (4), we find 
$f_4^*(6,1)=194.884$ and $f_5^*(6,1)=206.722$, with the corresponding 
multipliers $M_4(6)=-0.025$ and $M_5(6)=0.021$. From the minimal 
difference condition, we get $\tau=0.7182$, so that the exponential 
approximant (9) becomes $f_5^*(6)=201.692$, with the related multiplier 
$M_5^*(6)=0.021$. For the average approximant (11), we obtain 
$\brf_5(6)=200.803$, with the multiplier $\brM_5(6)=-0.004$.

The optimal self--similar exponential approximant is $\brf_5(6)=200.803$. 
The actual value of the index in {\it October 1929} was $194$, so that our 
forecast deviates from this by $3.51\%$. 

\section{Discussion}

We have shown that by means of the algebraic self--similar renormalization 
[8-10] it is possible to analyse stock market time series and even to 
predict market crashes. We have applied the approach to the time series 
corresponding to the {\it October 1997} crisis and to the {\it October 
1987} and {\it October 1929} stock market crashes. All these events, within
the self--similar renormalization procedure, are quite similar to each other.

For the 1987 and 1997 crises we have chosen for demonstration the 
dynamics of the NYSE Composite index. It is worth emphasizing that this 
choice is not principle, but is just a matter of convenience. The same 
analysis can be done for other representative indices. To show this, we 
present here such an analysis for the Nasdaq Composite index during the
{\it October 1997} crisis.

The dynamics of the latter index from April 30, 1997 to September 30, 
1997 has been as follows:
$$ 1260\;\; (Apr.\; 30), \qquad 1390\;\; (May\; 31), \qquad
1440\;\; (June\; 30), $$
$$ 1595\;\; (July\; 31), \qquad 1600\;\; (Aug.\; 31), \qquad 1690\;\;
(Sept.\; 30) . $$
Following the same way as above, we have in the three--point analysis 
$f_1^*(3,1)=1486,\; f_2^*(3,1)=1591$ and $M_1(3)=1,\; M_2(3)=-0.086$. 
Defining from (8) the effective time, we get for the approximant (9) the 
value $f_2^*(3)=1559$, with $|M_2^*(3)|=0.147$. For the averages (11) and 
(12), we find $\brf_2(3)=1538.5$ and $\brM_2(3)=0.457$.
Analogously, in the four--point analysis, we have $f_2^*(4,1)=1545,\;
f_3^*(4,1)=1911$ and $M_2(4)=-0.056,\; M_3(4)=0.148$. The self--similar
approximant (9) becomes $f_3^*(4)=1736$, with $|M_3^*(4)|=0.016$. And for 
(11) and (12), we get $\brf_3(4)=1728$, with $\brM_3(4)=0.046$. The 
five--point analysis yields $f_3^*(5,1)=1120,\; f_4^*(5,1)=1348$ and 
$M_3(5)=1.128,\; M_4(5)=-0.072$. Approximant (9) is $f_4^*(5)=1306$, with 
$|M_4^*(5)|=0.197$. Also, $\brf_4(5)=1234$, with $\brM_4(5)=0.528$. 
Finally, the six--point analysis gives $f_4^*(6,1)=1358,\; 
f_5^*(6,1)=1820$, with $M_4(6)=-0.01,\; M_5(6)=0.011$. Then, 
$f_5^*(6)=1623$, with $|M_5^*(6)|=0.008$ and $\brf_5(6)=1589$, with 
$\brM_5(6)=0.0005$.

Comparing all multipliers, we see that $\brM_5(6)$ has the minimal absolute
value. Therefore, the optimal forecast is $\brf_5(6)=1589$. The actual 
value of the Nasdaq Composite index on October 31, 1997 was $1593.61$. Our
forecast deviates from this by only $-0.29\%$.

Using for the given data (1) the polynomial representation (2), we obtain 
a set $\{ A_k\}$ of polynomial coefficients. The latter define the tendencies 
existing in the market, so that positive or negative coefficients correspond
to growth or decline, respectively. These tendencies compete with each 
other, analogously to heterophase fluctuations in statistical systems [19]. 
The state of a market at each time moment is presented as a superexponential 
function incorporating  the mixture of different tendencies. The optimal 
state is selected as the most stable one.

The possibility of treating market dynamics near a crisis as a self---similar 
evolution is based on the analogy between market crisis and critical phenomena.
The terms of a time series before a crisis contain a hidden information 
about this approaching crisis. The self--similar analysis plays the role 
of a decoder deciphering the hidden information.

The intensity of a crisis is determined by an interplay between the 
polynomial coefficients, whose positive or negative signs represent two 
competing tendencies, of growth or decay, respectively. The competition 
between these two different tendencies makes the market heterogeneous. The
heterogeneity of a market is a necessary condition for the occurrence of 
a crisis, which happens when one of two competing tendencies becomes 
dominant. This is in complete analogy with phase transitions in heterophase 
statistical systems [19].

It is worth noting that the suggested approach can be applied only to 
those markets whose evolution is governed by the collective behaviour of 
interacting agents. Such markets can be called self--regulated or 
self--organized. It is only these markets are analogous to complex 
statistical systems whose behaviour is caused by internal reasons. In the 
case of some strong external forces acting on a market, it cannot be 
considered as self--regulated and, consequently, it looses the 
possibility of being described by self--similar dynamics. If the external 
influence is not too strong, a market can be rather stochastic at a 
short--time scale but on average self--similar at a longer time scale 
[7]. We plan to give a more detailed consideration of these questions in 
forthcoming publications.

\end{document}